\def\eeq{\end{equation}}

\def\beq{\begin{equation}}

\def\bea{\begin{eqnarray}}

\def\eea{\end{eqnarray}}

\documentstyle[aps]{revtex}

\begin{document}

\title{Circular-like Maps: Sensitivity to the Initial Conditions,
Multifractality and Nonextensivity}

\author{U\v{g}ur T{\i}rnakl{\i}$^1$,
Constantino Tsallis$^2$, Marcelo L. Lyra$^3$}

\address{
$^1$ Department of Physics, Faculty of Science, Ege University,
35100 Izmir-Turkey\\
$^2$ Centro Brasileiro de Pesquisas Fisicas, Rua Xavier Sigaud 150,
22290-180 Rio de Janeiro - RJ, Brazil\\
$^3$ Departamento de Fisica, Universidade Federal de Alagoas, 57072-970
Maceio-AL, Brazil\\
e-mails: tirnakli@sci.ege.edu.tr, tsallis@cat.cbpf.br, marcelo@fis.ufal.br}

\maketitle

\begin{abstract}
Dissipative one-dimensional maps may exhibit special points
(e.g., chaos threshold) at which the Liapunov exponent vanishes. Consistently,
the sensitivity to the initial conditions has a power-law time dependence,
instead of the usual exponential one.
The associated exponent can be identified with $1/(1-q)$, where $q$
characterizes the nonextensivity of a generalized entropic form currently used
to extend standard, Boltzmann-Gibbs statistical mechanics in order to cover a
variety of anomalous situations.
It has been recently proposed [Lyra and Tsallis, Phys. Rev. Lett. {\bf 80}
(1998) 53] for such maps the scaling law
$1/(1-q)=1/\alpha_{min} - 1/\alpha_{max}$, where
$\alpha_{min}$ and $\alpha_{max}$ are the extreme values appearing in the
multifractal $f(\alpha)$ function. We generalize herein the usual circular map
by considering inflexions of arbitrary power $z$, and verify that the scaling 
law holds for a large range of $z$. Since, for this family of maps,
the Hausdorff dimension $d_f$ equals unity $\forall z$ in contrast with
$q$ which does depend on $z$, it becomes clear that $d_f$ plays no major
role in the sensitivity to the initial conditions.

\noindent
{\it PACS Number(s): 05.45.+b, 05.20.-y, 05.70.Ce}

\end{abstract}

\newpage


\section{Introduction}
Whenever a physical system has long-range interactions and/or long-range
microscopic memory and/or evolves in a (multi)fractal-like space-time, the
extensive, Boltzmann-Gibbs (BG) statistics might turn
out to be inadequate in the sense that it fails
to provide {\it finite} values for relevant thermodynamical quantities of the
system. In order to theoretically deal with nonextensive systems of this 
(or analogous) kind, two major formalisms are available: the so-called
quantum groups \cite{qg} and the generalized
thermostatistics (GT), proposed by one of us a decade
ago \cite{tsallis1}. This two formalisms present in fact deep
connections \cite{abe}. We focus here the GT.
Within this framework, nonextensivity is defined through a generalized
entropic form, namely
\beq
S_q=k\frac{1-\sum_{i=1}^W p_i^q}{q-1}\;\;\;\;\;(q \in {\cal R})
\eeq
where $k$ is a positive constant and $\{p_i\}$ is a set of probabilities
associated to $W$ microscopic configurations. We can immediately check that
the $q \rightarrow 1$ limit recovers the usual, extensive, BG entropy
$-\sum_{i=1}^W p_i\;ln\;p_i$. Also, if a composed system $A+B$ has
probabilities which factorize into those corresponding to the subsystems
$A$ and $B$, then $S_q(A+B)/k=S_q(A)/k+S_q(B)/k+(1-q)S_q (A)S_q(B)/k^2$.
This property exhibits the fact that $q$ characterizes the degree of
nonextensivity of the system.

Especially  during the last five years, a wealth of works have appeared
within this formalism. In fact, it is possible to (losely) classify these
works as follows: (i) Some of them \cite{concepts}
address the generalization of  relevant concepts and properties of
standard thermostatistics, such as Boltzmann's H-theorem,
fluctuation-dissipation theorem, Onsager reciprocity theorem, among
others; (ii) Other GT works \cite{verify} focuse applications to some
physical systems where BG statistics is known to fail (stellar
polytropes, turbulence in electron-plasma, solar neutrino problem,
peculiar velocities of spiral galaxies, Levy anomalous diffusion, among
others), and yields satisfactory results; (iii) Finally, an area of
interest which is progressing rapidly, addresses the long standing puzzle
of better understanding the physical meaning of the entropic index $q$.
This line concerns the study of nonlinear dynamical systems (both low
\cite{TPZ,costa,marcelo,marcelo2} and high \cite{tamarit,papa} dimensional
dissipative ones, as well as Hamiltonian systems \cite{celia}) in
order to clarify the connection between $q$, the sensitivity to the
initial conditions and a possible (multi)fractality hidden in the
dynamics of the system. This paper belongs to the last class of efforts
and is organized as follows. In Section 2 we briefly summarize recent
related results. In Section 3 we introduce a new map which generalizes
the circular one, and study its main properties. Finally, we conclude in
Section 4.

\section{Power-law Sensitivity to Initial Conditions}
The most important dynamical quantities that are used to characterize the
chaotic systems are the Liapunov exponent $\lambda_1$ and the
Kolmogorov-Sinai entropy $K_1$ (the meaning of the subindex 1 will soon
become transparent).  Let us define, for a one-dimensional map of the
real variable $x$, the quantity  $\xi (t) \equiv \lim_{\Delta
x(0)\rightarrow 0} \frac{\Delta x(t)}{\Delta x(0)}$ , where  $\Delta
x(0)$ and $\Delta x(t)$ are discrepancies of the initial conditions at
times $0$ and $t$ respectively. It can be shown that, under quite generic
conditions, $\xi$ satisfies the differential equation
$d\xi/dt=\lambda_1\;\xi$, hence $\xi(t)=\exp(\lambda_1t)$.
Consequently, if $\lambda_1<0$ ($\lambda_1>0$) the system is said to be
{\it strongly} insensitive (sensitive) to the initial conditions.
Similarly, for a dynamical system under certain conditions, we can define
$K_1$ as essentially the increase, per unit time, of
$S_1\equiv -\sum_{i=1}^W p_i\;ln\;p_i$. Furthermore, it can be shown
that, with some restrictions, $K_1=\lambda_1$ if $\lambda_1 \ge 0$ and
$K_1=0$ otherwise. This is frequently referred to as the Pesin equality
\cite{hilborn}.

The case we are focusing in the present work is the so called {\it
marginal} case, corresponding to $\lambda_1=0$.  It has been argued
\cite{TPZ,costa,marcelo} that, in this case, the differential equation
satisfied by $\xi$ is  $d\xi/dt=\lambda_q\;\xi^q$, hence

\beq
\xi (t)  = \left[1+(1-q)\lambda_q t\right]^{1/(1-q)} ,
\eeq

\noindent
where $\lambda_q$ is the generalized Liapunov exponent. One can verify
that $q=1$ recovers the standard, {\it exponential} case whereas $q\neq
1$  yields a {\it power-law} behavior. If $q>1$ ($q<1$) the system is
said to be {\it weakly} insensitive (sensitive) to the initial
conditions. Furthermore, for this marginal case we can define the
generalized Kolmogorov-Sinai entropy $K_q$ as the increase, per unit
time, of $S_q$. Finally, for this anomalous case, it has been argued
\cite{TPZ} that the Pesin equality itself can be generalized as follows:
$K_q=\lambda_q$ if $\lambda_q \ge 0$ and $K_q=0$ otherwise.

Recently, these ideas have been applied to some dissipative
one-dimensional maps (a logistic-like and a periodic-like map, sharing
the same universality class, as well as the standard circular map, which
belongs to a different universality class), and the numerical results
suggested a close relationship between the nonextensivity parameter $q$
and the fractal (Hausdorff) dimension $d_f$ associated with the dynamical
attractor \cite{costa,marcelo,marcelo2}. Very specifically, in those examples,
when $d_f$ approaches unity (which is the Euclidean dimension of the
system) from below then $q$ also approaches unity from below, and does
that in a {\it monotonic manner}. Naturally, this fact strongly suggests
that the validity of the statistical $q=1$ (BG) picture is intimately
related to the {\it full} occupancy of the phase space. However, an
important question which remains open is whether the full occupancy
is {\it sufficient} for having a BG scenario. We will show here that it
is not! We shall introduce in this paper a generalized circular map
(characterized by an inflexion power $z$), and show that, at the critical
point, $d_f(z)=1\;(\forall z)$, and nevertheless $q<1$ !  This fact
consistently complements the new scaling relation proposed in
\cite{marcelo}, namely
\beq
\frac{1}{1-q} = \frac{1}{\alpha_{min}} - \frac{1}{\alpha_{max}} \;\; ,
\eeq

\noindent  where $\alpha_{min}$ and $\alpha_{max}$ are the extremes of
the multifractal singularity spectrum $f(\alpha)$ of the attractor (for
details see \cite{halsey}). This relation clearly indicates that once the
scaling properties of the dynamical attractor are known, one can precisely
infer the proper entropic index $q$ that must be used  for other purposes.

\section{A Family of Circular-like Maps}
The circle map is an iterative mapping of one point on a circle to
another of the same circle. This map describes dynamical systems
possessing a natural
frequency $\omega_1$ which are driven by an external force of frequency
$\omega_2$; $\Omega\equiv \omega_1 / \omega_2$  is known as the "bare"
winding number. These systems tend to mode-lock at a frequency
$\omega_1^*$ and $\omega \equiv \omega_1^* / \omega_2$  is known as the
"dressed" winding number.  The standard circle map, for one-dimension, is
given by

\beq
\theta_{t+1} = \Omega + \theta_t - \frac{K}{2\pi} \sin(2\pi\theta_t)
~~~~~ mod(1),
\eeq

\noindent   with $0<\Omega<1$ ; $0<K<\infty$.
For $K<1$ the circle map is linear at the vicinity of its extremal 
point and  exhibits only periodic motion.  From now on we take $K=1$, 
the onset value above which chaotic orbits
exist. For this
map, once mode-locked, 
$\omega = \lim_{t\rightarrow\infty} (\theta_{t+1}-\theta_t)$  
remains constant and rational for a small range
of the parameter $\Omega$ with the "dressed" versus "bare" winding 
number curve exhibiting a "devil staircase" aspect\cite{jensen}
 (if $\theta_{t+1} < \theta_t$ then one shall use  
$\omega = 1+\theta_{t+1}-\theta_t$ in order to leave it $mod(1)$). 
At the onset to chaos, a set of zero measure
and universal scaling dynamics is produced at special irrational dressed
winding numbers which have the form of an infinite continued-fraction 
expansion \begin{equation}
\omega=\frac{1}{n+\frac{1}{m+\frac{1}{p+...}}} ,
\end{equation}
with   $n,m,p,..$ integers. The best studied one is when $\omega$ equals the
golden mean, i.e., $\omega_{GM} = (\sqrt5 -1)/2$ 
($n=m=p=... =1$)\cite{shenker,feigenbaum}. It is worth mentioning that the 
golden mean is the asymptotic ratio between consecutive numbers of the 
Fibonacci series ($\lim_{n\rightarrow\infty}F_n/F_{n+1} = (\sqrt{5}-1)/2$, 
where $F_0 = 0$, $F_1 = 1$ and $F_n=F_{n-1}+F_{n-2}$). In order to determine 
the bare winding number at this critical point, we iterate the map for a 
large number of time steps ($10^6$ steps) starting with $\theta_0=0$ and use 
a linear regression to numericaly compute $\omega$. The bare winding number 
$\Omega$ is then adjusted to have the renormalized winding number 
$\omega$ equal to the golden mean, which results in $\Omega_c=0.606661...\;$ 
With these parameters, the standard circle map has a
cubic inflexion ($z=3$) near its extremal point ${\bar \theta}=0$. 

A generalized version of the circle map can be defined as

\beq
\theta_{t+1} = \Omega + \left[\theta_t - \frac{1}{2\pi}
\sin (2\pi\theta_t)\right]^{z/3} ~~~,
\eeq

\noindent   where $z>0$ ($z=3$  reproduces the standard case). For
every value of $z$, the golden mean of $\omega$ corresponds to different
"bare" winding numbers, which we call as $\Omega_c$. In order to determine
these critical values of $\Omega$, one searches, within a given precision
($12$ digits in our calculations), the value of $\Omega$ corresponding to 
$\omega_{GM}$ with the same precision. The calculated values
of $\Omega_c$, as a function of $z$, are shown in Fig. 1. The numerical
values are indicated in the Table. We remark that, in the limit
$z\rightarrow 0$ ($z\rightarrow\infty$) we verify that $\Omega_c\propto
z^{1/2}$ ($1-\Omega_c\propto 1/z^{\beta}$ with $\beta\simeq 0.41$).

For our present purpose, a very important feature of this map is that,
for every value of $z$, the critical attractor visits the {\it entire}
circle ($0<\theta_t$ (mod 1) $<1$)
and therefore has a support fractal dimension $d_f = 1$. The calculated
values of $d_f$, using a box counting algorithm, are indicated in Fig. 2.
In what concerns the sensitivity to the initial conditions, the
function $\xi (t)$ is given by

\bea
\ln \xi(t)= \ln\left|\frac{d\theta_N}{d\theta_0}\right| =
\sum_{t=1}^{N} \ln\left\{\frac{d}{d\theta}\left[\theta_t -\frac{1}{2\pi}
\sin(2\pi\theta_t)\right]^{z/3}\right\}= \nonumber \\
\;\;\;\;\;\;\;\;\;\;\; =\sum_{t=1}^N \ln
\left\{\frac{z}{3}\left[\theta_t - \frac{1}{2\pi}
\sin(2\pi\theta_t)\right]^{\frac{z}{3}-1}
\left[1-\cos(2\pi\theta_t)\right]\right\}
\eea

\noindent  and displays, for $\Omega =\Omega_c$, a power-law divergence,
$\xi\propto t^{1/(1-q)}$,
from where the value of $q$ can be calculated 
by measuring, on a log-log plot, the upper bound slope, $1/(1-q)$. 
In Fig. 3, the
$z=4.5$ and $z=6$ cases have been illustrated; see also the Table.

In order to check the accuracy of  the scaling
relation (3), one needs to determine  the $\alpha_{min}$ and $\alpha_{max}$
values of the $f(\alpha )$ curve. Therefore, one has to study the 
structure of the trajectory $\theta_1$, $\theta_2$,...,$\theta_i$,... and 
to estimate the singularity spectrum (strength of singularities $\alpha$ and 
their  fractal dimensions $f$) of this Cantor-like set. To perform the 
numerical calculation we truncate the series $\theta_i$ at a chosen 
Fibonacci number $F_n$ (we recall that $F_n/F_{n+1}$ gives the golden mean 
and therefore defines the proper  scaling factor). The distances $l_i$ 
between consecutive points of the set define the natural scales for the 
partition with measures $p_i=1/F_n$ attributed to each segment. 
After that, the singularity spectrum can be directly obtained following a 
standard prescription\cite{halsey}. In general,
$\sup_{\alpha} f(\alpha)=d_f$ and, in the present case,
$f(\alpha_{min})=f(\alpha_{max})=0$. We define $\alpha_{top}$ through 
$f(\alpha_{top})=d_f$. 

However, the situation for the generalized
circle map is somewhat different than that of the standard circle map in
the sense that the standard one has a fast convergence of the 
$f(\alpha )$ curve when larger number of iterations are considered,  
whereas the generalized map presents only a slow and oscillatory convergence. 
An example of a sequence of $f(\alpha)$ curves obtained from increasing 
number of iterations is shown in Figure 4. Notice its non-monotonic behavior, 
specially near the upper edge. 
In Figure 5, we plot the numerically obtained values of $\alpha_{min}$, 
$\alpha_{top}$ and 
$\alpha_{max}$ as a function of $1/\ln{N}$, where $N$ is the number of 
iterations. From these data, we are not able to accuratelly estimate  
their asymptotic values for large map inflexion $z$. In Figure 6 we plot
 the extrapolated $f(\alpha )$ curves for typical values of $z$. Although 
these show the main expected trends of the singularity spectra, namely 
$z$-dependent shape but $d_f=1$ for all $z$, their extremal points may
 need further corrections.  
We shall point out that this feature is inherent to the numerical method used 
to estimate  the $f(\alpha )$ curve representing the singularity strengths of 
a multifractal measure. Its extremal points are governed by the scaling 
behavior of the most concentrated ($\alpha_{min}$) and most rarefied 
($\alpha_{max}$) sets in the measure, the latter being usually poorly 
sampled.

An alternative method for computing the extremal values of the singularity 
strengths $\alpha$ can be obtained by studying how the distances around 
$\theta = 0$ scale down as the trajectory $\theta_i$ is truncated at two 
consecutive Fibonacci numbers, $F_n$, $F_{n+1}$. Shenker has found that this 
distance shall scale by a universal factor $\alpha_F(z)$\cite{shenker} 
($F$ stands for Feigenbaum). 
This region around $\theta =0$ corresponds to the most rarefied one so that 
$l_{-\infty}\sim [\alpha_F(z)]^{-n}$. The corresponding measure scales as 
$p_{-\infty}=p_i= 1/F_n\sim (\omega_{GM})^n$, which leads 
to\cite{marcelo,halsey,feigenbaum2}
\begin{equation}
\alpha_{max} = \frac{\ln{p_{-\infty}}}{\ln{l_{-\infty}}} = 
\frac{\ln{ \omega_{GM}}}{\ln{[\alpha_F(z)]^{-1}}}~~~.
\end{equation}
Following along the same lines, the most concentrated region on the set 
shall scale down as $\l_{+\infty}\sim\alpha_F(z)^{-zn}$ while 
$p_{+\infty}\sim [\omega_{GM}]^n$, so that
\begin{equation}
\alpha_{min} = \frac{\ln{p_{+\infty}}}{\ln{l_{+\infty}}} = 
\frac{\ln{\omega_{GM}}}{\ln{[\alpha_F(z)]^{-z}}}~~~.
\end{equation}
Eqs. (8) and (9) imply
\begin{equation}
\alpha_{max}/\alpha_{min}=z
\end{equation}

In Figure 7, we show the critical sequence of the distances around 
$\theta =0$ for a large value of the map inflexion ($z=8$). The data 
provide an accurate estimation of the universal factor 
$\alpha_F(z=8) = 1.1568$ from which precise values of $\alpha_{min}$ and 
$\alpha_{max}$ can be inferred. In the Table we list the results for $3<z<8$, 
together with  the values of $\alpha_{top}$ from the extrapolated 
$f(\alpha )$ curves and $q$ 
obtained from both the scaling relation (3) and from the sensitivity 
function (7). 
Notice that the scaling relation (3) 
is satisfied for all $z$ (see also Figure 8), just like the case of the 
logistic-like maps \cite{marcelo}.
Furthermore, these results indicate two other important points:
(i) Even though the fractal dimension of the support $d_f$ is Euclidean 
(i.e., $d_f=1$) for 
all $z$, the system sensitivity to initial conditions is still 
$z$-dependent. 
(ii) What matters is $\alpha_{min}$ and $\alpha_{max}$, and not $d_f$, 
in other words, what precisely controls the entropic index $q$ is not $d_f$ 
but the sensitivity 
to initial conditions, reflected by the multifractal nature of the attractor.

\section{Conclusions}
In this paper, we contribute to the field of  low-dimensional dissipative 
systems by introducing a convenient generalization (with inflexion power $z$) 
of the standard circular map and then studying its critical point (analogous 
to the chaos threshold of logistic-like maps). More precisely, we have 
numerically studied its sensitivity to the initial conditions and have shown 
that it is given by a {\it power}-law ($\xi \propto t^{1/(1-q)}$) instead of 
the usual {\it exponential} behavior. 
Moreover, we have shown that this example, as the logistic-like maps, 
satisfies the scaling law given in Eq. (3). Although $q,\;\alpha_{min}$ and 
$\alpha_{max}$ depend on $z$, $d_f$ {\it does not}.
This is a quite important result because it illustrates that, for having a 
Boltzmann-Gibbs scenario ($q=1$), it is not enough to fully occupy the phase 
space during the dynamical evolution of the system. What is essentially 
necessary is to have a {\it quick}, exponential-like occupation of the phase 
space, so that ergodicity and mixing are naturally attained. 

In addition to this, it is worth mentioning that there are also other
efforts along this line which address  high-dimensional dissipative
systems, namely those exhibiting self-organized
criticality \cite{SOC}. Amongst them, the study of the Bak-Sneppen model for 
biological evolution \cite{tamarit} and the Suzuki-Kaneko model for the 
battle of birds defending their territories \cite{papa} can be enumerated. 
In both cases it is shown that, at the self-organized critical state, 
a power-law sensitivity to initial conditions emerges, 
like in the present case.

As a final remark, it should be emphasized that these ideas seem to be 
valid and applicable not only to low- and high-dimensional dissipative 
systems but also to conservative (Hamiltonian) systems with long-range 
interactions . This fact has been illustrated very recently \cite{celia} 
on the long-range classical $XY$ ferromagnetic model, whose entire Liapunov 
spectrum collapses (for an infinitely wide energy interval) to zero at the 
thermodynamic limit if (and only if) the 
range of the interactions is sufficiently long.  Further efforts focusing, 
along these lines, both dissipative and conservative systems
are welcome.

\subsection*{Acknowledgments}
One of us (UT) is a TUBITAK M\"{u}nir Birsel Foundation Fellow and
acknowledges its financial support which made his visit to CBPF (Rio de
Janeiro) possible. He would also like to thank CBPF for
kind hospitality during his stay. This work was partially supported by
PRONEX and CNPq (Brazilian Agencies).



\newpage

{\bf Table and Figure Captions}

\vspace{1.5cm}

Table : Our best numerical values for $\Omega_c$, $\alpha_F$, $\alpha_{min}$, 
$\alpha_{top}$, $\alpha_{max}$ and  $q$ from both the scaling relation (3) 
and from the sensitivity function (7) for various $z$. 
It is worthy to mention that we numerically verify that, 
for $z \ge 3$,  $\alpha_{top} \ge [\alpha_{min}\; \alpha_{max}]^{1/2}$ 
(the equality appears to hold for $z=3$). 

\vspace{1cm}

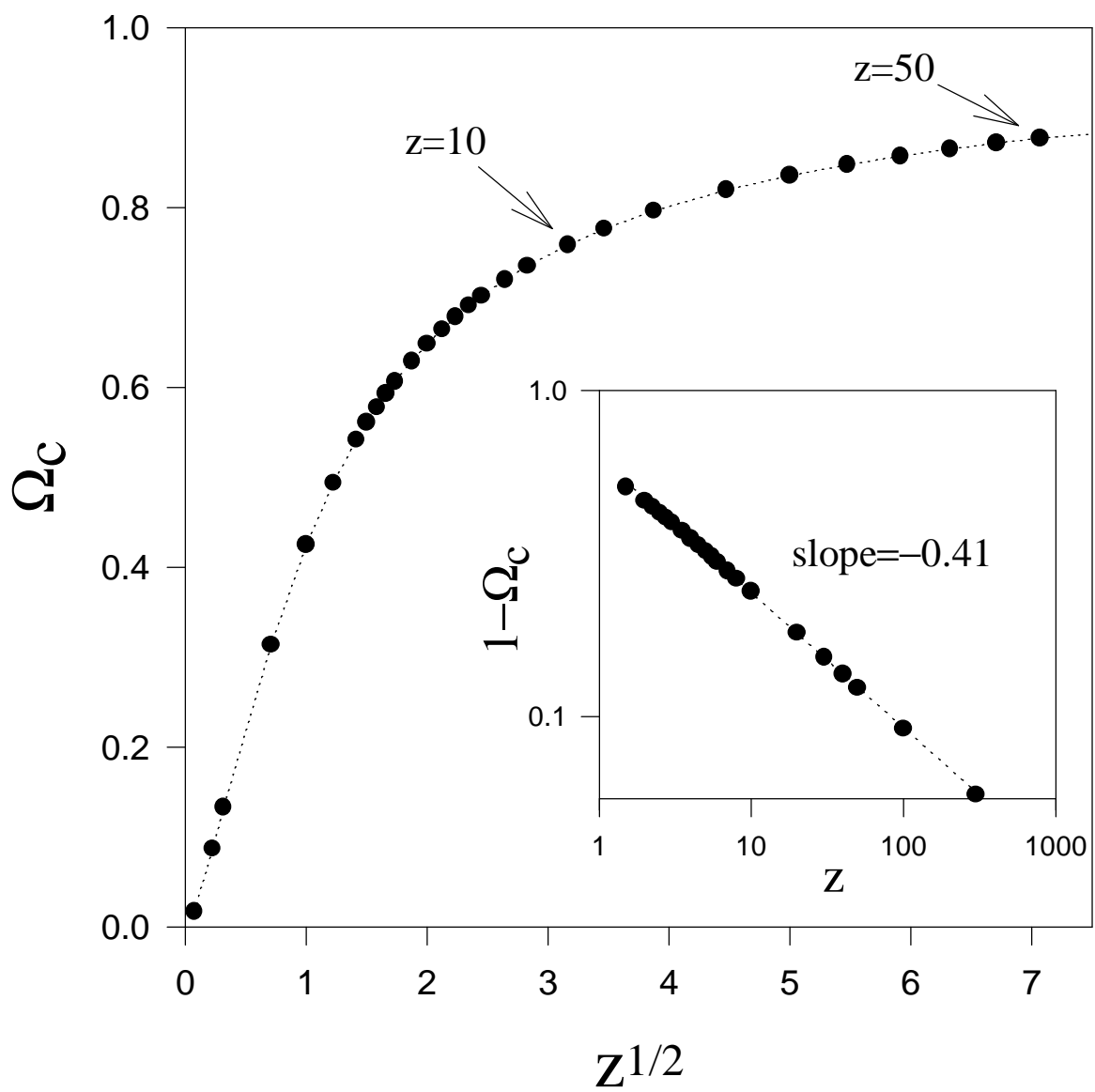
Figure 1 : The values of the "bare" winding number $\Omega_c$ as a 
function of $z$ for the generalized circle maps.

\vspace{1cm}

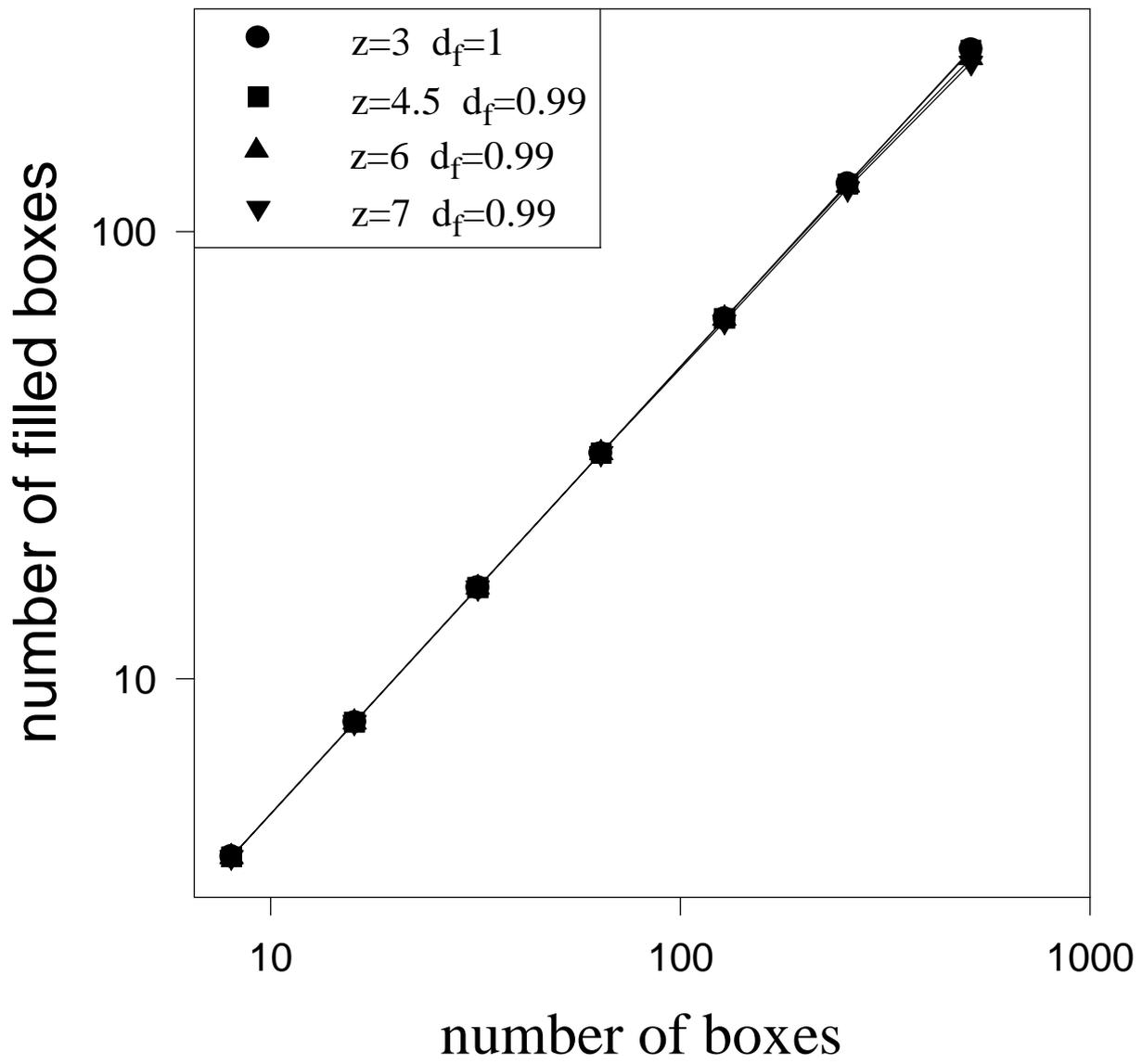
Figure 2 : Box counting graph for determining the fractal dimension 
$d_f$ of the critical attractor for typical values of $z$.

\vspace{1cm}

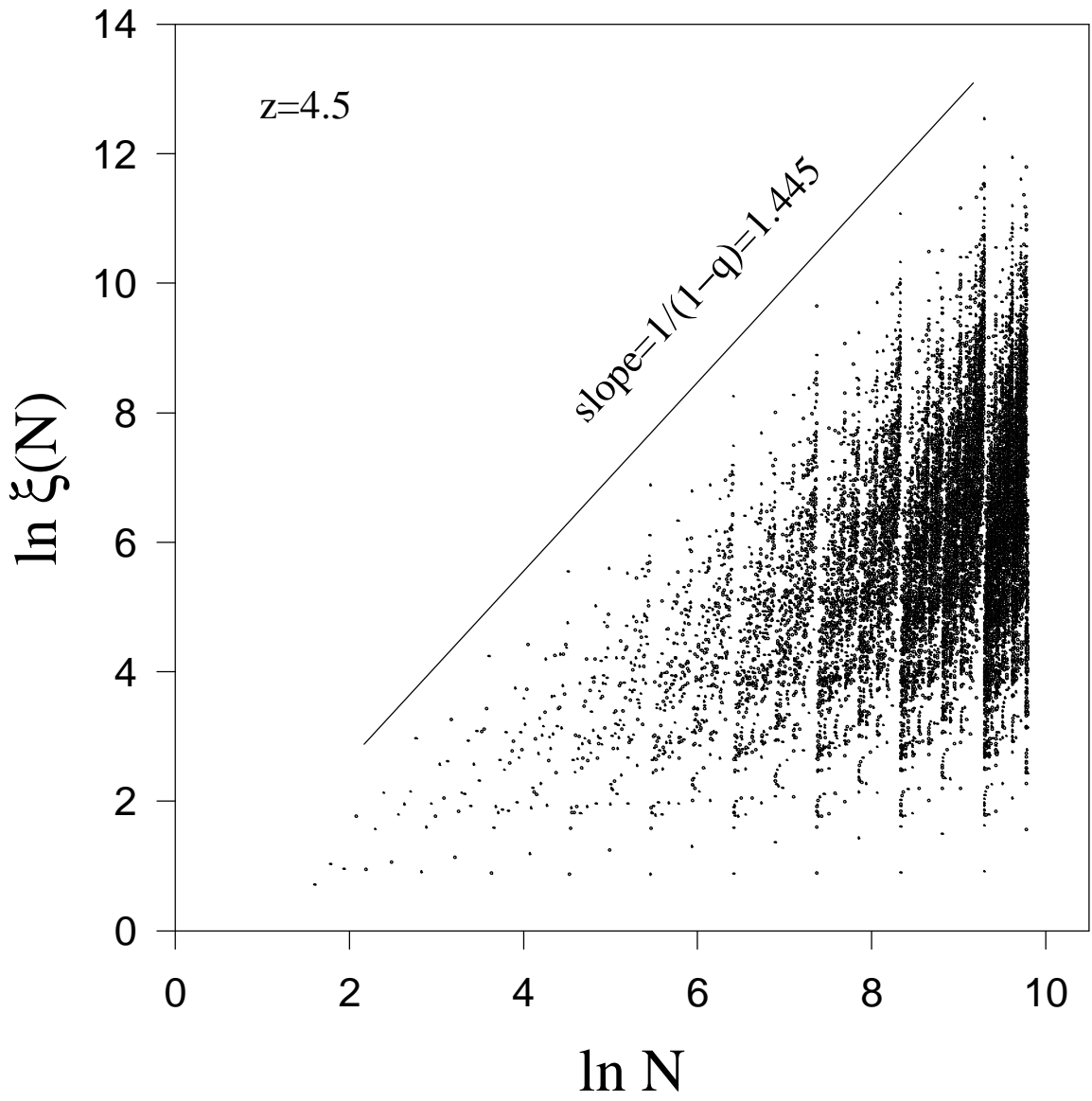
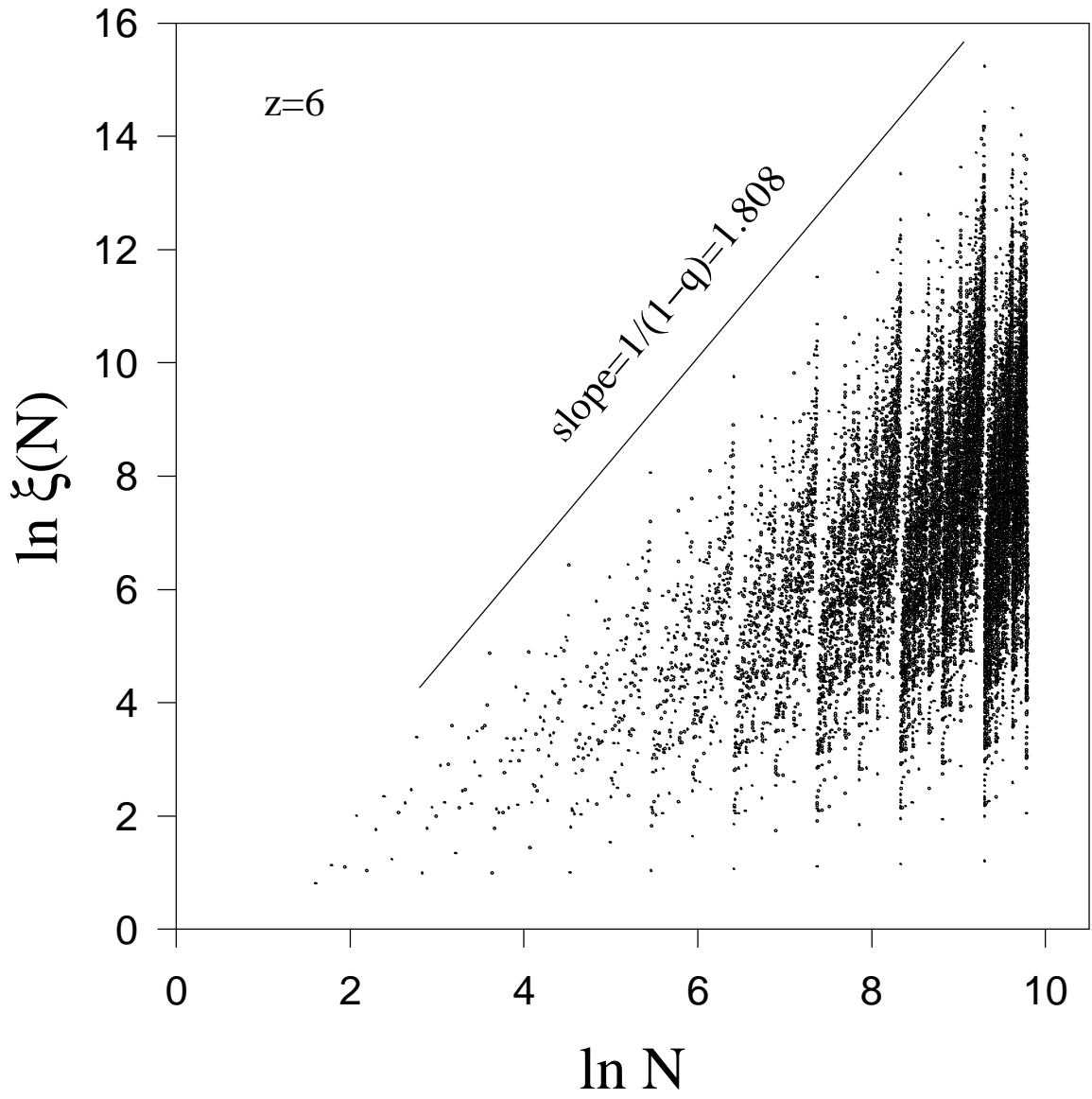
Figure 3 : The plot of $\ln \xi(N)$ versus $\ln N$. (a) for $z=4.5$ and
(b) for $z=6$.

\vspace{1cm}

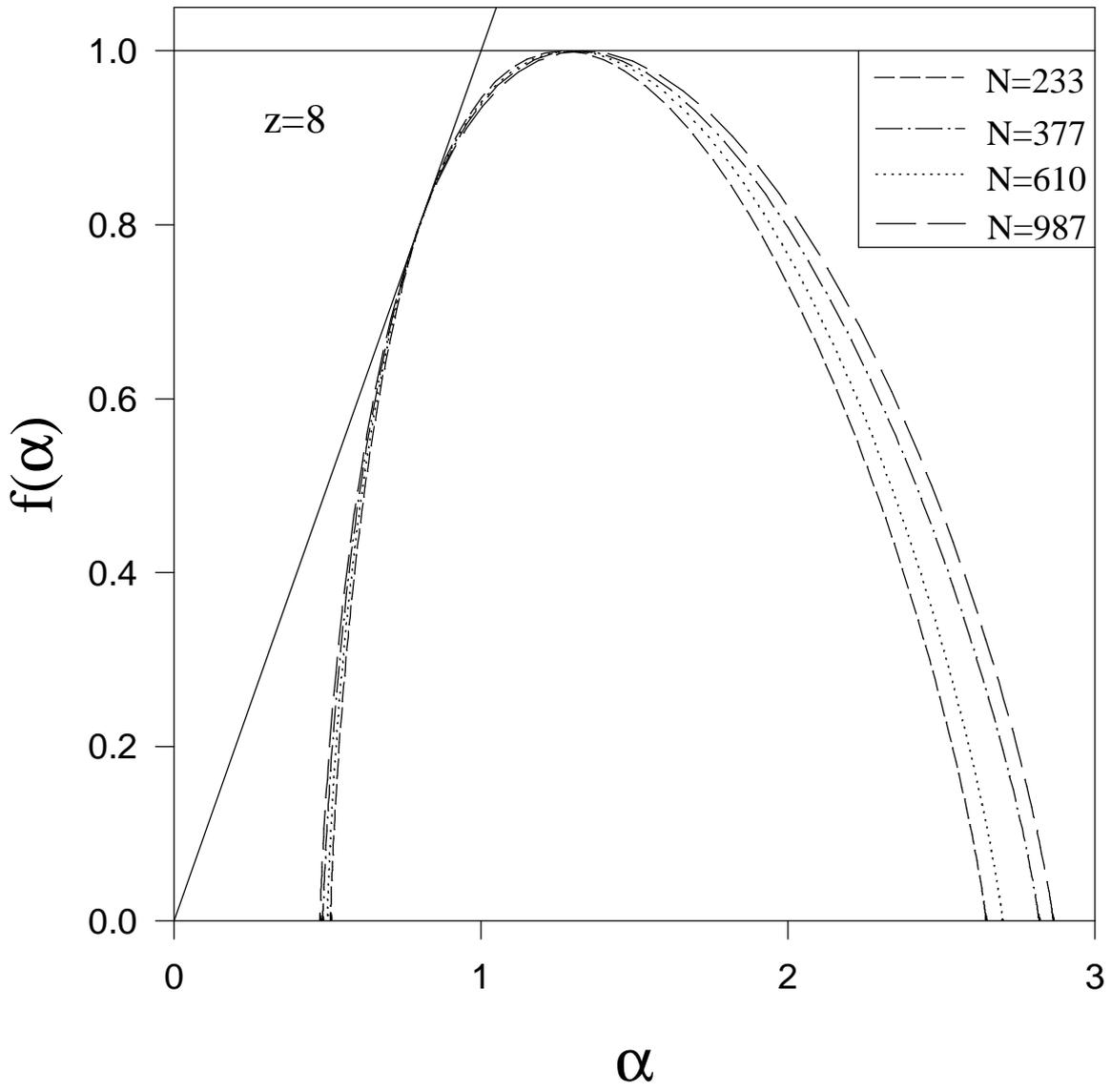
Figure 4 : Approximate multifractal singularity spectra of the critical 
attractor of the $z=8$ generalized circular map for increasingly large 
number of iterations, going from $N=233$ to $987$. 

\vspace{1cm}

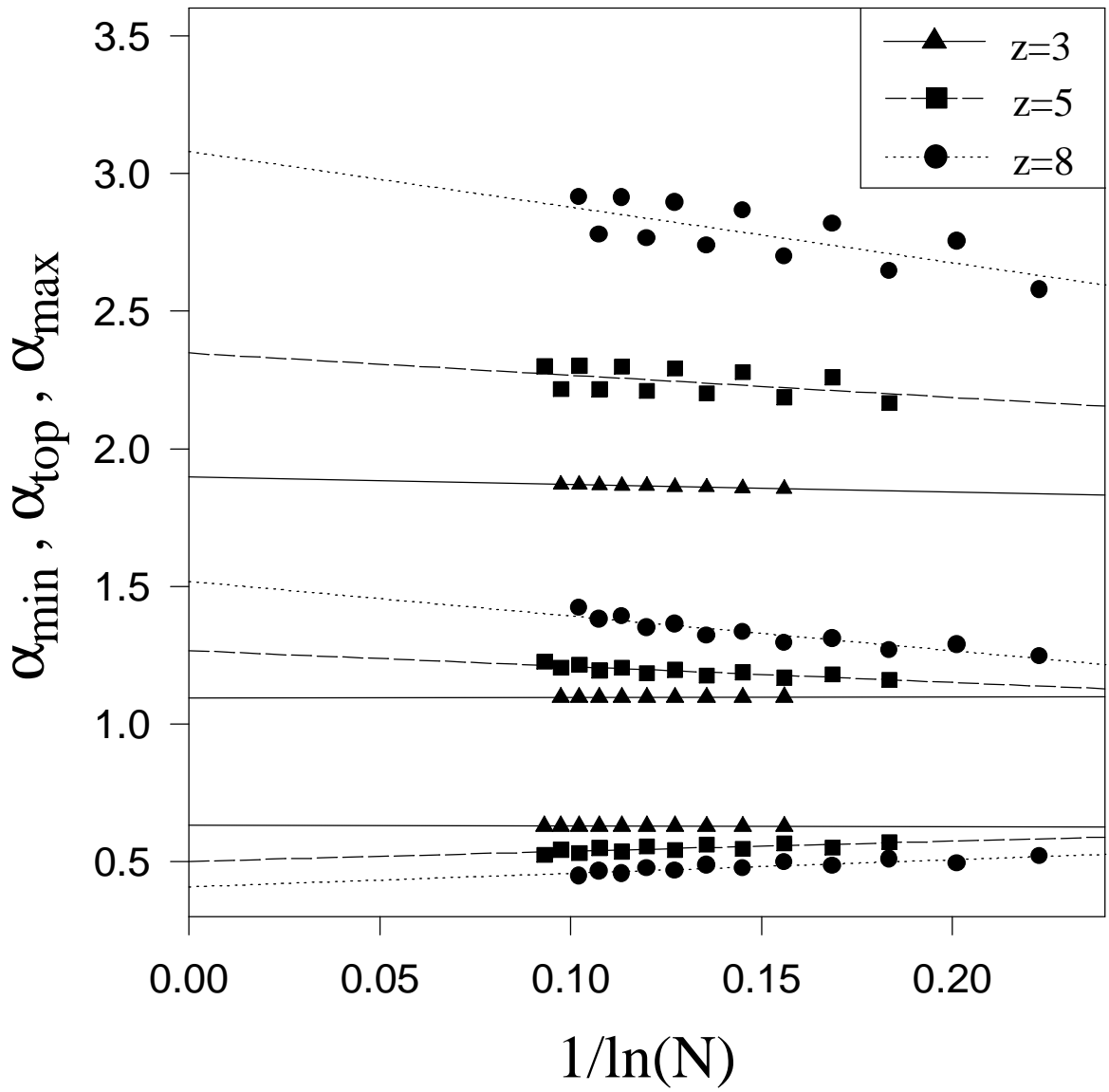
Figure 5 :  $\alpha_{min}$ (bottom three lines), $\alpha_{top}$ 
(middle three lines) and $\alpha_{max}$ (top three lines) from the 
singularity spectra obtained from distinct  number of iteractions and 
typical values of $z$.  Notice the slow and oscillatory convergence.

\vspace{1cm}

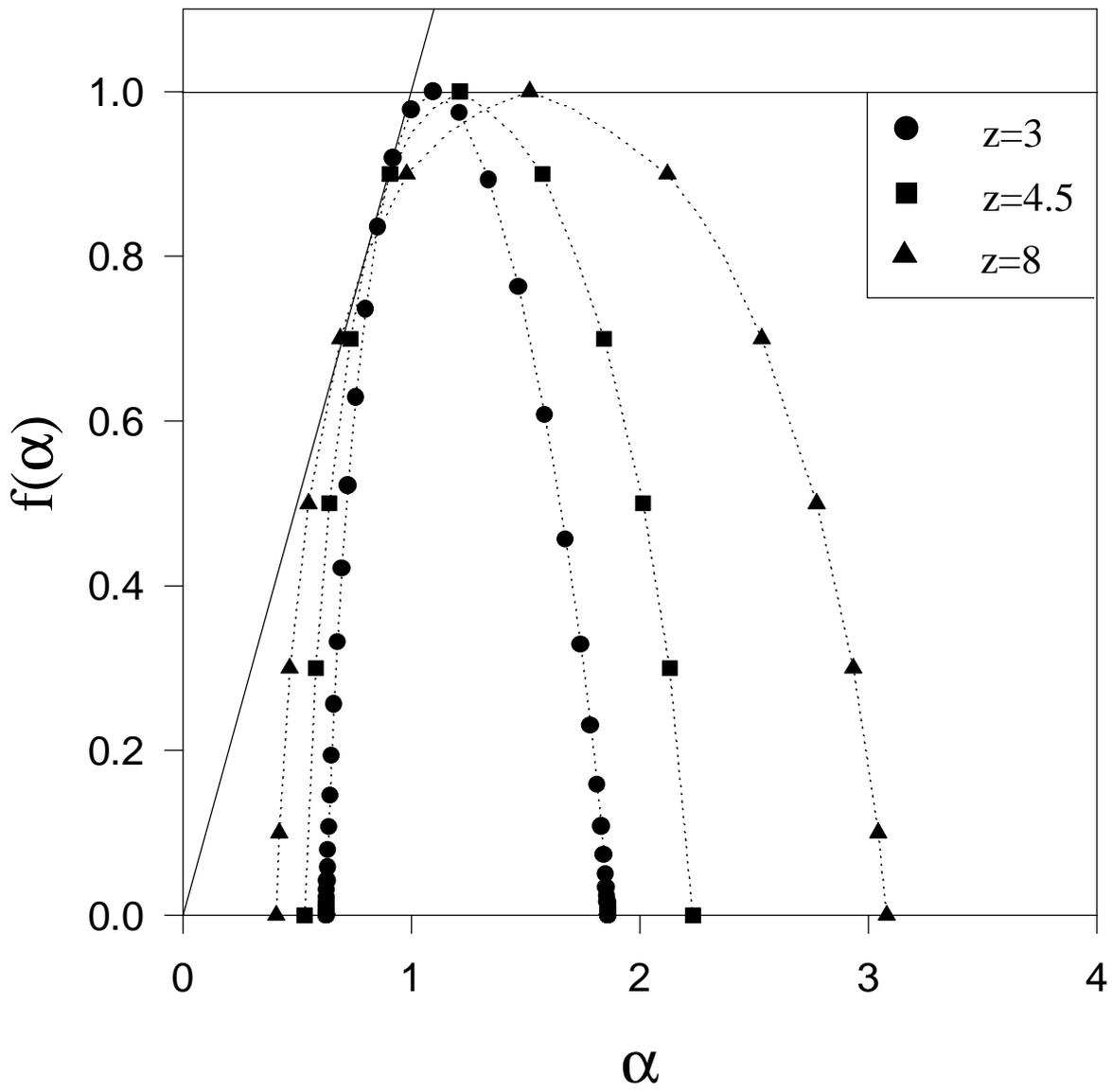
Figure 6 : The extrapolated $f(\alpha )$ curves for typical values of 
the map inflexion $z$. Notice that, although the shape is $z$-dependent,
 they present $d_f = 1$ for all $z$. The solid lines correspond to 
$f(\alpha)=1$ and $f(\alpha)=\alpha$. The dotted lines are guides to the eye.

\vspace{1cm}

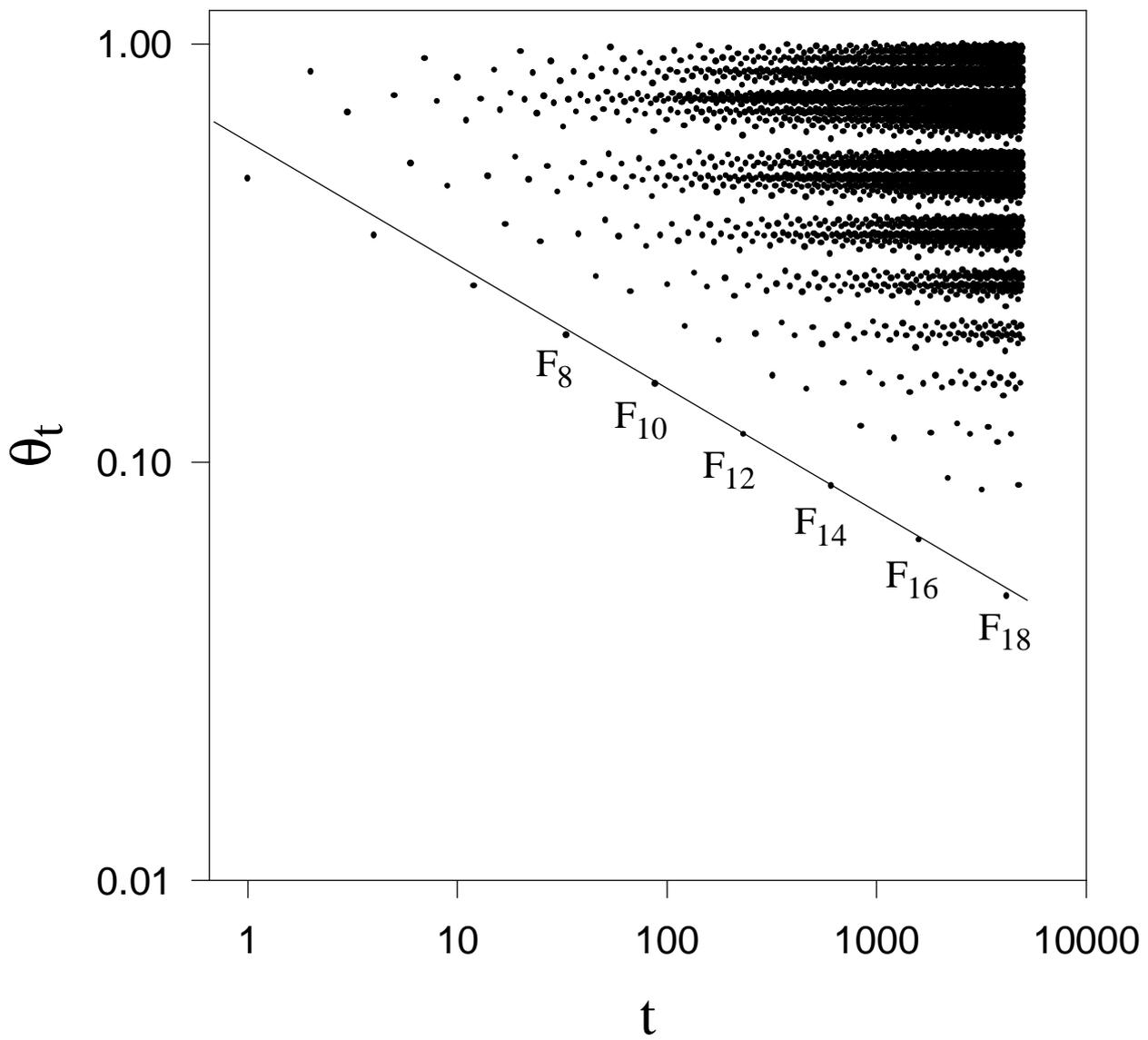
Figure 7 : The sequence $\theta_t$ as a function of $t$ for $z=8$. 
The minimal distance to $\theta =0$ scales down as $t^{-0.305}$. 
Using that  $F_n\sim\omega_{GM}^{-n}$ for large $n$ one obtains that 
$\l_{-\infty}\sim\alpha_F^{-n}$, with $\alpha_F(z=8) = 1.158$. 

\vspace{1cm}

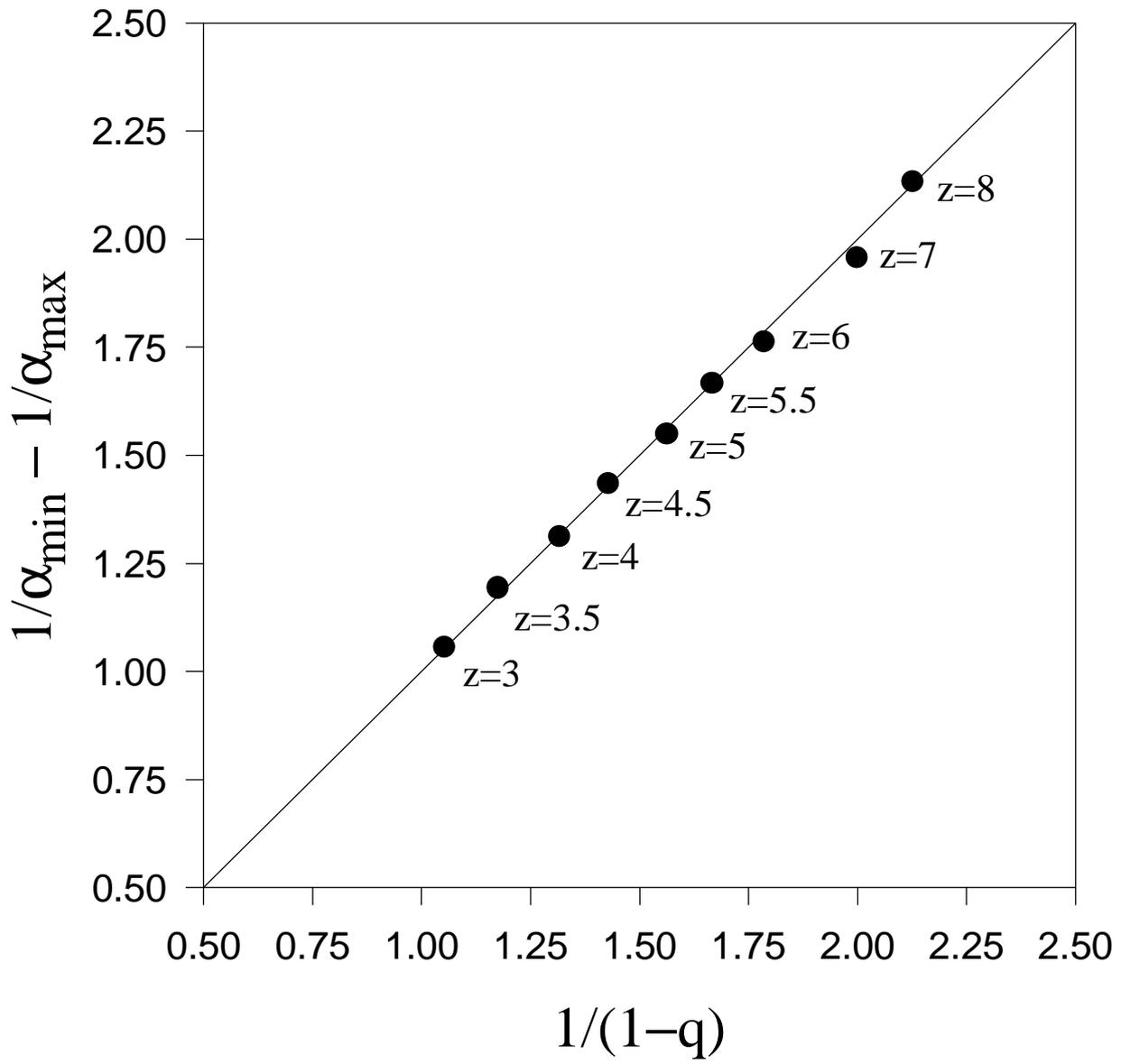
Figure 8 : $1/\alpha_{min}-1/\alpha_{max}$  versus $1/(1-q)$ values from the 
sensitivity function. The straight line represents the scaling prediction 
(Eq. (3)).

\newpage

\begin{center}
{\bf Table}

\vspace{2cm}

\begin{tabular}{||c|c|c|c|c|c|c|c||}  \hline
$z$  & $\Omega_c$ & $\alpha_F$ & $\alpha_{min}$ & $\alpha_{top}$  &
$\alpha_{max}$  &  $q$ & $q$ \\
       &                      &           &         &       &        &
  (Eq.(3))     & (Eq.(7))  \\
\hline
$\;3.0\;$  & $\;0.606661063469...$ &$1.289$    &$0.632$ & $1.096 $  & $1.895$ & 
$0.05\pm 0.01$ & $0.05\pm 0.01$  \\ \hline
$\;3.5\;$  & $\;0.629593799039...$ &$1.258$    &$0.599$ & $1.124 $  & $2.097$  &
$0.16\pm 0.01$ & $0.15\pm 0.01$  \\ \hline
$\;4.0\;$  & $\;0.648669091983...$ &$1.234$    &$0.572$ & $1.167 $  & $2.289$  &
$0.24\pm 0.01$ & $0.24\pm 0.01$  \\ \hline
$\;4.5\;$  & $\;0.664861001064...$ &$1.218$    &$0.542$ & $1.213 $  & $2.440$  &
$0.30\pm 0.01$ & $0.30\pm 0.01$  \\ \hline
$\;5.0\;$  & $\;0.678831756505...$ &$1.205$    &$0.516$ & $1.266 $  & $2.581$  &
$0.36\pm 0.01$ & $0.36\pm 0.01$  \\ \hline
$\;5.5\;$  & $\;0.691048981515...$ &$1.195$    &$0.491$ & $1.314 $  & $2.701$  &
$0.40\pm 0.01$ & $0.40\pm 0.01$  \\ \hline
$\;6.0\;$  & $\;0.701853340894...$ &$1.185$    &$0.473$ & $1.351 $  & $2.838$  &
$0.43\pm 0.01$ & $0.44\pm 0.01$  \\ \hline
$\;7.0\;$  & $\;0.720182442561...$ &$1.170$    &$0.438$ & $1.451 $  & $3.065$  &
$0.49\pm 0.01$ & $0.50\pm 0.01$  \\ \hline
$\;8.0\;$  & $\;0.735233625356...$ &$1.158$    &$0.410$ & $1.518 $  & $3.280$  &
$0.53\pm 0.01$ & $0.53\pm 0.01$  \\ \hline
\end{tabular}

\end{center}

\end{document}